\documentclass[preprintnumbers,showpacs,amsmath,amssymb,prd,floatfix,twocolumn,nofootinbib,superscriptaddress]{revtex4}
\usepackage{graphicx}
\usepackage{epsfig}
\usepackage{bm}
\usepackage{amsfonts}

\usepackage{epsfig}
\usepackage{bm}

\begin{document}

\title{Unified dark energy thermodynamics: varying $w$ and the $-1$-crossing}

\author{Emmanuel N.~Saridakis} \affiliation{Department of
Physics, University of Athens, GR-15771 Athens, Greece}

\author{Pedro F. Gonz\'{a}lez-D\'\i az}
\affiliation{Colina de los Chopos, Instituto de F\'\i sica
Fundamental, Consejo Superior de Investigaciones Cient\'\i ficas,
Serrano 121, 28006, Madrid, Spain}

\author{Carmen L. Sig\"{u}enza}
\affiliation{Estaci\'{o}n Ecol\'{o}gica de Biocosmolog\'{i}a,
Pedro de Alvarado 14, 06411 Medell\'{i}n, Spain}

\begin{abstract}
We investigate, in a unified and general way, the thermodynamic
properties of dark energy with an arbitrary, varying
equation-of-state parameter $w(a)$. We find that all quantities
are well defined and regular for every $w(a)$, including at the
$-1$-crossing, with the temperature being negative in the phantom
regime ($w(a)<-1$) and positive in the quintessence one
($w(a)>-1$). The density and entropy are always positive while the
chemical potential can be arbitrary. At the $-1$-crossing, both
temperature and chemical potential are zero. The temperature
negativity can only be interpreted in the quantum framework. The
regular behavior of all quantities at the $-1$-crossing, leads to
the conclusion that such a crossing does not correspond to a phase
transition, but rather to a smooth cross-over.
\end{abstract}

 \maketitle

\section{Introduction}\label{intro}

Recent cosmological observations support that the universe is
experiencing an accelerated expansion, and that the transition to
the accelerated phase has been realized in the recent cosmological
past \cite{observ}. In order to explain this remarkable behavior,
and despite the intuition that this can be achieved only through a
fundamental theory of nature, physicists can still propose some
paradigms for its description, such are theories of modified
gravity \cite{ordishov}, or ``field'' models of dark energy.  The
field models that have been discussed widely in the literature
consider a canonical scalar field (quintessence) \cite{quint}, a
phantom field, that is a scalar field with a negative sign of the
kinetic term \cite{phant}, or the combination of quintessence and
phantom in a unified model named quintom \cite{quintom}. The
common feature of all the field models of dark energy, is that the
equation-of-state parameter $w$ of the dark energy ``fluid'', is
induced by the field evolution and thus it acquires a dynamical
nature ($w\rightarrow w(a)$ with $a$ the scale factor).

On the other hand, there have been large efforts in order to study
the thermodynamic properties of dark energy fluids with a
{\it{constant}} $w=w_0\neq-1$. For the case of quintessence
scenarios, where $w=w_0>-1$, one can use the results of
conventional perfect fluid thermodynamics. However, in the case of
phantom fields, where $w=w_0<-1$, a thermodynamic investigation
must be performed from first lines. Some authors assume a zero
chemical potential $\mu$ and argue that the temperature of a
phantom fluid must be negative, with the density and the entropy
positive \cite{phantomtherm,Myung:1900km}, and the negative nature
of the temperature is related to the quantum properties of phantom
fields (although in some initial works on $\mu=0$ the entropy was
found to be negative and thus the phantom phase meaningless
\cite{Lima:2004wf}). On the other hand, some authors consider the
non-zero chemical potential case, and argue that phantom fluids
must have $\mu<0$, with temperature, entropy and density being
positive, with the phantom particles possessing a bosonic,
massless nature \cite{Lima:2008uk,Pereira:2008af}.

In our opinion, the central point of dark energy models is not so
much that $w$ can take a constant value different from $-1$, but
rather that $w(a)$ is varying (indeed since $w(a)$ is a ratio of
two varying quantities, being a constant requires rather special
constructions). In addition, the crossing of the phantom-divide
$w=-1$ from above to below, is not only possible but it could
indeed be the basic requirement for a successful description of
observations. Therefore, the question of investigating in a
unified way the thermodynamics of dark energy fluids with  an
arbitrarily varying $w(a)$, becomes crucial.

However, a comment must be made here, concerning the applicability
of thermodynamics to time-dependent systems such as the universe,
where gravitational degrees of freedom are present. Although in
principle statistical mechanics assumes equilibrium and
ergodicity, which could be violated in time-dependent
gravitational backgrounds, and despite the fact that a
thermodynamic and statistical description of the dynamics of
gravitational degrees of freedom does not fully exist yet, we
believe that such an approach can be enlightening. This is similar
to the thermodynamical investigation of the properties of black
holes or de Sitter spaces, that is purely gravitational systems
evolving in time. In addition, it is fairly known that black holes
can be regarded as the time-reversed analog of the Big Bang
universe, despite starting from static spacetime metrics. Having
these points in mind we proceed to the thermodynamic analysis of a
general dark-energy fluid. The paper can be outlined as follows:
In Sec. \ref{genthrm} we present the thermodynamic properties of a
general-$w(a)$ dark-energy fluid, while  Sec. \ref{cosmiml} is
dedicated to discussion, concluding remarks and a brief summary of
the results.

\section{General $w(a)$ Thermodynamics}\label{genthrm}

We consider dark-energy fluids, with energy density and pressure
respectively $\rho(a)$ and $p(a)$, and with an arbitrary and
general equation of state of the form:
\begin{equation}\label{eos}
p(a)=w(a)\rho(a).
\end{equation}
In these expressions $a$ is the scale factor of a homogenous and
isotropic Friedmann-Robertson-Walker universe, with metric:
\begin{equation}\label{metric}
ds^{2}=dt^{2}-a^{2}(t)\left(\frac{dr^2}{1-k
r^2}+r^2d\theta^2+r^2\sin^2\theta \,d\phi^2\right),
\end{equation}
where $k=0,\pm1$ is the curvature parameter and $t$ is the
comoving time. Since we desire our analysis to be general, we do
not consider specific cosmological models, with explicitly
extracted Friedmann equations, but we just consider the resulting
$w(a)$. Thus, the present investigation is valid for every FRW
cosmological model. Finally, note that we could equivalently
express $w(a)$ as $w(t)$ or $w(z)$ (with $z$ the redshift), since
such a transformation is straightforward.

As usual, the equilibrium thermodynamic states of a relativistic
perfect fluid are characterized by an energy-momentum tensor
$T^{\mu\nu}$:
\begin{equation}\label{energymom}
T^{\mu\nu}=(p+\rho)u^\mu u^\nu-p g^{\mu\nu},
\end{equation}
where $u^{\mu}$ ($\mu =\{0,1,2,3\}$) is the  4-velocity in the
metric $g^{\mu\nu}$. The particle and entropy currents, $N^\mu$
and $S^\mu$ respectively, are defined as:
\begin{equation}\label{NScurrents}
N^\mu=n u^\mu,\, \ \ S^\mu=s u^\mu,
\end{equation}
with $n$ and $s$  the densities of particle number and entropy.
Denoting  the covariant derivative by $(;)$, the equations of
motion are given by $T^{\mu\nu};_\nu=0$, $N^\mu;_\mu$ and
$S^\mu;_\mu=0$, which in the case of the FRW geometry write:
 \begin{eqnarray}\label{eoms}
 \dot{\rho} + 3 [1 + w(a)]\rho \frac{\dot a}{a}&=&0\\
 \dot{n} + 3n\frac{\dot a}{a}&=&0\\
 \dot{s} + 3s\frac{\dot
 a}{a}&=&0,
\end{eqnarray}
with the dot denoting the comoving time derivative. In the case of
a general $w(a)$, the solution for $\rho(a)$ reads: {\small{
\begin{eqnarray}\label{solwvar}
&&\rho(a)=\rho_0\,\exp\left\{\int_a^{a_0}\frac{3[w(a)+1]}{a}da\right\}\Rightarrow\nonumber\\
&&\rho(a)=\rho_0\left\{\frac{a_0^{3[w_0+1]}}{a^{3[w(a)+1]}}\right\}\,\exp\left[-3\int_a^{a_0}da\,w'(a)\ln
a\right],
\end{eqnarray}}}
where the prime denotes the $a$-derivative, and $a_0$ and
$w_0\equiv w(a_0)$ are the present values of the scale factor and
of the dark energy equation-of-state parameter respectively. In
this relation we preferred to use $w'(a)$, in order for the
extraction of the $w=const$ results (i.e. $w'(a)\rightarrow0$) to
be straightforward. Similarly, the solutions for $n(a)$ and $s(a)$
are simply:
\begin{equation}\label{NSsol}
n(a)=n_0 \left(\frac{a_0}{a}\right)^{3},\, \ \ s(a)=s_0
\left(\frac{a_0}{a}\right)^{3},
\end{equation}
with $n_0$ and $s_0$ the values of the corresponding quantities at
present (from now on the index $0$ stands for the present value of
a quantity). Finally, we stress that $w(a)$ is the value of the
equation-of-state parameter when the scale factor is $a$, while
$w_0$ is its present value. Obviously, on can face all the
evolution combinations, where $w(a)$ can cross $-1$ at a specific
scale factor or not, and/or $w_0$ being $-1$ or not, or the case
$w(a)=w_0=const.$ with the constant being not equal or equal to
$-1$ (the last case is the cosmological constant universe).
Therefore, in the following one must examine the regularity of all
expressions when $w(a)$ and/or $w_0$ is equal to $-1$.

Let us now refer to thermodynamics. Considering $(n,T)$ as the 2D
thermodynamical base, we use the identity (Gibbs law): $T
\biggl({\partial p \over \partial T}\biggr)_{n}=\rho + p - n
\biggl({\partial \rho \over \partial n}\biggr)_{T}$, which in the
case of a perfect fluid writes:
\begin{equation}\label{Gibbs2}
T \biggl[{\partial\, w(a)\rho(a) \over \partial
T}\biggr]_{n}=[w(a)+1]\rho(a) - n(a) \biggl[{\partial \rho(a)
\over
\partial n}\biggr]_{T}.
\end{equation}
 On the other hand, we can express $\dot{\rho}$ as $
\dot{\rho}=\dot{n}\left(\frac{\partial \rho}{\partial
n}\right)_T+\dot{T}\left(\frac{\partial \rho}{\partial
T}\right)_n$. Combining this relation with (\ref{Gibbs2})  and
(\ref{eoms}) we acquire $\left[3\frac{\dot{a}}{a}\,
T(a)w(a)+\dot{T}(a)\right]\left[\frac{\partial \rho(a)}{\partial
a}\right]_n=-3\frac{\dot{a}}{a}\,T(a)\rho(a)w'(a)$, or using
(\ref{solwvar}):
\begin{equation}\label{eqT2}
\left[\frac{3}{a}\,
T(a)w(a)+T'(a)\right]\left[w(a)+1\right]=T(a)w'(a).
\end{equation}
Note that this general differential equation leads
straightforwardly to the general physical implication that $T(a)$
must be zero when $w(a)$ crosses $-1$. Furthermore, note that in
the case where $w(a)\equiv-1=const.$ at all times, i.e. of a pure
cosmological constant, equation (\ref{eqT2}) is trivially
satisfied, not allowing for the determination of $T(a)$, and this
has led some authors to argue that the temperature of the vacuum
state is ill-defined \cite{Myung:1900km}. However, we see that
considering a general $w(a)$, i.e with  $w'(a)\neq0$, we are led
to $T=0$ for the vacuum state, which is a self-consistent result.

If $w(a)\neq-1$, the general solution of (\ref{eqT2}) is:{\small{
\begin{equation}\label{Tsol}
T(a)=T_0\,\left[\frac{w(a)+1}{w_0+1}\right]\left[\frac{a_0^{3w_0}}{a^{3w(a)}}\right]\,\exp\left[-3\int_a^{a_0}da\,w'(a)\ln
a\right],
\end{equation}}}
with $T_0$ the present temperature value. Solution (\ref{Tsol})
holds for $w(a)<-1$ or $w(a)>-1$. For $w(a)=-1$ the only
information that we can obtain straightaway from (\ref{eqT2}) is
that $T(a)$ must be zero. However, this requirement is indeed
satisfied by (\ref{Tsol}), and furthermore we can see that:
\begin{equation}\label{Tlim}
\lim_{w(a)\rightarrow-1^\pm}T(a)=0.
\end{equation}
Thus, we can safely consider that (\ref{Tsol}) is the general
solution for every $w(a)$, including the $-1$-crossing. We mention
that if $w_0\rightarrow-1$, then as required by (\ref{eqT2})
$T_0\rightarrow0$ too, and $T(a)$ remains regular.

Expression (\ref{Tsol}) has an interesting and general physical
implication, namely that  $T(a)$ and $w(a)+1$ (and thus $T_0$ and
$w_0+1$) have always  the same sign. Therefore, in summary, the
temperature is always negative for $w(a)<-1$, it is zero for the
cosmological constant bound, and it is always positive for
$w(a)>-1$. Later on, we will discuss these points in detail.
Finally note that we could denote the regular ratio $T_0/(w_0+1)$
by a positive constant $C_0$, but we prefer keeping it in order
for the various quantities at present to be straightforwardly
obtained.

A useful relation can be obtained using (\ref{solwvar}) and
(\ref{Tsol}), namely:
\begin{equation}\label{rhoTT}
\rho(a)=\rho_0\left\{\frac{T(a)}{T_0}\frac{[w_0+1]}{[w(a)+1]}\right\}\left(\frac{a_0}{a}\right)^{3},
\end{equation}
which is regular at $w(a)=-1$ (and at $w_0=-1$), as can be seen
from (\ref{Tsol}). Equivalently, eliminating $a^3$ between
(\ref{Tsol}) and (\ref{rhoTT}) we acquire the generalized
Stefan-Boltzmann law:
\begin{eqnarray}\label{rhoTTT}
\rho(T)=\rho_0\left\{\frac{T(a)}{T_0}\frac{[w_0+1]}{[w(a)+1]}\right\}^{\frac{w(a)+1}{w(a)}} a_0^{\frac{3[w(a)-w_0]}{w(a)}}\cdot\   \ \ \ \nonumber\\
 \ \ \ \ \ \ \cdot\exp\left[\frac{3}{w(a)}\int_a^{a_0}da\,w'(a)\ln
a \right].\
\end{eqnarray}
Note that in the case $w(a)=0$, i.e when the dark energy fluid is
a dust, one can only use  relation (\ref{rhoTT}), since the
elimination of $a^3$  between (\ref{Tsol}) and (\ref{rhoTT}) is
not possible.

Let us now return to thermodynamics, using the general identity
$Ts=\rho+p-\mu n$, with $\mu$ the chemical potential of the dark
energy fluid, considered non-zero in general. This relation can be
used to determine $\mu$. In particular, written as
\begin{equation}\label{Tsmu}
\mu(a)=\frac{[w(a)+1]\rho(a)}{n(a)}-\frac{T(a)s(a)}{n(a)},
\end{equation}
 and using (\ref{solwvar}) and (\ref{Tsol})
it leads to:{\small{
\begin{equation}\label{mua}
\mu(a)=\mu_0\left[\frac{w(a)+1}{w_0+1}\right]\left[\frac{a_0^{3w_0}}{a^{3w(a)}}\right]\exp\left[-3\int_a^{a_0}da\,w'(a)\ln
a \right],
\end{equation}}}
where we have defined
\begin{equation}\label{mu0}
\mu_0\equiv\frac{\rho_0(w_0+1)-T_0s_0}{n_0}
\end{equation}
the chemical potential value at present.
 Expressions (\ref{mua})
and (\ref{mu0}) are regular at $w_0=-1$ (since in this case
$T_0=0$ too), and furthermore at $w(a)=-1$ (\ref{mua}) leads to
$\mu(a)=0$. Note however that in general the sign of $\mu_0$, and
thus of $\mu(a)$, can be arbitrary. Therefore, we conclude that
the chemical potential for the cosmological constant bound is
zero, but it is arbitrary at the two sides of that bound,
depending on the specific cosmological model. Later on, we will
discuss these points in detail.  Finally, (\ref{Tsmu}) with the
use of (\ref{rhoTT}), leads to the simple relation:
\begin{equation}\label{muT}
\mu(T)=\mu_0\left[\frac{T(a)}{T_0}\right].
\end{equation}

Relation (\ref{Tsmu}), with $s(a)=S(a)/V(a)$, can be used to
calculate the total entropy $S$ of the universe, with physical
volume $V(a)=a^3$. Indeed, using (\ref{NSsol}), (\ref{rhoTTT}) and
(\ref{muT}), we acquire:
\begin{eqnarray}\label{S2}
S(T)=s_0V(a)\left\{\frac{T(a)}{T_0}\frac{[w_0+1]}{[w(a)+1]}\right\}^{\frac{1}{w(a)}}a_0^{\frac{3[w(a)-w_0]}{w(a)}}\cdot\  \nonumber\\
\cdot\exp\left[\frac{3}{w(a)}\int_a^{a_0}da\,w'(a)\ln
 a\right],\ \ \
\end{eqnarray}
which is regular and non-zero at $w(a)=-1$. It can be easily seen,
using (\ref{rhoTT}), that
\begin{equation}\label{Stot}
S(a)=s_0 V(a)\left(\frac{a_0}{a}\right)^3=s_0V_0=s(a)V(a),
\end{equation}
with $V_0=a_0^3$ the current physical volume of the universe.
Expression (\ref{Stot}) is just the statement of the conservation
of entropy in the whole universe, and provides a self-consistency
test for our calculations. Finally, note that in the case
$w(a)=0$, i.e when the dark energy fluid is a dust, relations
(\ref{S2}) and (\ref{rhoTTT}) lead to a trivial result, and one
can use only relation (\ref{Stot}).

In the aforementioned investigation we desired to remain general,
and we did not use any ansatz for $w(a)$. However,
phenomenologically, one can consider various
$w(z)$-parametrizations \cite{Szydlowski:2008zzaB} at will, in
order to extract quantitative predictions for the quantities at
hand. Lastly, taking the limit $w'(a)\rightarrow0$ (or
equivalently just setting $w'(a)=0$ since the terms that need
caution have been written as separated pre-factors), we extract
the corresponding relations for the $w(a)=w_0=const$ case:
\begin{eqnarray}
&&\rho(a)=\rho_0\left(\frac{a_0}{a}\right)^{3[w_0+1]}\nonumber\\
&&T(a)=T_0\left(\frac{a_0}{a}\right)^{3w_0}\nonumber\\
&&\mu(a)=\mu_0\left(\frac{a_0}{a}\right)^{3w_0}\nonumber\\
&&\rho(T)=\rho_0\left\{\frac{T(a)}{T_0}\right\}^{\frac{w_0+1}{w_0}}\nonumber\\
&&\mu(T)=\mu_0\left[\frac{T(a)}{T_0}\right]\nonumber\\
&&S(T)=s_0V(a)\left\{\frac{T(a)}{T_0}\right\}^{\frac{1}{w_0}}.
\label{oldrelations}
\end{eqnarray}
We mention that these relations are valid also for the simple
cosmological constant ($w_0=-1$ and $T_0=\mu_0=0$), giving
$\rho=\rho_0=const.$, $T=\mu=0$ and $S=s_0V_0=s(a)V(a)=const.$.

\section{Discussion}\label{cosmiml}

Let us discuss about the physical context of the obtained results.
First of all, we see that all quantities are regular and
well-defined, for all values of $w(a)$. In addition, the density
and entropy are always positive, consistently with the basic
requirements of classical and quantum field theory. This behavior
clarifies some ambiguities about the phantom nature, since one
does not need integrability conditions \cite{Myung:1900km} or
special constructions
\cite{Lima:2004wf,Lima:2008uk,Pereira:2008af} in order  to make
this ``phase'' possible (which remarkably might indeed be the
current phase of the universe).

An important physical consequence is that the temperature of a
dark energy fluid with $w(a)<-1$ is negative, independently of the
value of the chemical potential which can be arbitrary. This is in
contrast with the works that required necessarily  a negative
$\mu$ and a positive $T$ for the constant-$w$ phantom fluids
\cite{Lima:2008uk,Pereira:2008af}. The misleading point in these
considerations were that since $w(a)\equiv w_0=const$, the authors
instead of equation  (\ref{eqT2}) used $ 3 T(a)w(a)+aT'(a)=0$,
which leads to qualitative different results. The reason for this
behavior is that we face a singular perturbation problem
\cite{Hilbert}, i.e the results for $w'(a)\rightarrow0$ are more
general and do not coincide in the whole variable-range with those
for $w(a)\equiv w_0=const.$. Furthermore, note that if one
considers $w(a)\equiv w_0=const.$ then he cannot educe any result
at all for the temperature sign, since this will coincide with the
sign of $T_0$, which is arbitrary in principle. Only allowing for
a $-1$-crossing (i.e for a general $w(a)$) we can safely conclude
that $T(a)$ and $w(a)+1$ (and thus $T_0$ and $w_0+1$) have the
same sign, as can be seen from (\ref{Tsol}).

Another significant physical implication, is that the vacuum,
either as a permanent state ($w(a)\equiv w_0=-1$ at all times) or
as an instantaneous state at the time of the crossing of a
varying-$w(a)$, has zero temperature and zero chemical potential.
This was expected, since we can consider it to correspond to the
zero mode of the various involved scalar fields. On the contrary,
a vacuum with non-zero $T$ and $\mu$ would require a non-trivial
explanation. We mention that starting with $w(a)\equiv w_0=-1$,
one cannot find a relation for the temperature (and thus for the
chemical potential too), and therefore he concludes that the
temperature and chemical potential for this state is ambiguous or
ill-defined \cite{Myung:1900km}, or he is led to wrong results
\cite{Lima:2008uk,Pereira:2008af}. The correct approach is to
start with a general $w(a)$ and then examine the limit
$w(a)\rightarrow-1$.

The regular behavior of all quantities at the phantom-divide
crossing, leads to an additional interesting result. In
particular, we conclude that such a crossing does not correspond
to a phase transition, but rather to a cross-over. An additional
argument for this statement arises from the regular behavior of
the ``specific heat'', defined as $\propto\frac{\partial
\rho}{\partial T}$ or $\propto T\frac{\partial S}{\partial T}$, at
the crossing. The $-1$-crossing, dynamically is just a pass to the
super-accelerated evolution, while thermodynamically is just a
smooth pass from $T>0$ to $T<0$, without a phase transition. This
behavior was observationally required, since if a radical
cosmological phase transition had taken place in the recent
cosmological past, it would have left observable imprints.

The temperature negativity can only be interpreted in the quantum
framework. Although the discussion about the construction of
quantum field theory of phantoms is still open in the literature
(see for example \cite{Cline:2003gs} about  causality and
stability problems and the possible spontaneous breakdown of the
vacuum into phantoms and conventional particles), there have been
serious attempts in overcoming these difficulties and construct a
phantom theory consistent with the basic requirements of quantum
field theory \cite{quantumphantom}. Thus, as it was analyzed in
detail in \cite{phantomtherm}, in such a quantum consideration of
the phantom fluid all  novel phenomena (stimulated absorption of
phantom energy, generalized Wien and Planck radiation laws) can be
naturally positioned.

One of these novel phenomena, that can have interesting
cosmological implications, is the accretion of dark energy onto
black holes. In particular, in \cite{Babichev:2004yx} it was shown
that the solution for a stationary, spherically symmetric,
accretion of a cosmological perfect fluid onto a Schwarzschild
black hole with mass $M$, is given by $\dot{M}=4\pi A M^2
(\rho+p)$, with $A$ a positive constant, which in the case of a
general $w(a)$ reads:
\begin{equation}
\dot{M}=4\pi A M^2 [1+w(a)]\rho(a)\label{Mdot}.
\end{equation}
Although a specific solution of this equation requires a constant
$w(a)$ (see \cite{phantomtherm}), in general we observe that
$\dot{M}>0$ in the quintessence regime, $\dot{M}<0$ in the phantom
one, while $\dot{M}=0$ at the $-1$-crossing, consistently with the
generalized second law of thermodynamics (the Hawking radiation is
neglected in this consideration). Relation (\ref{Mdot}) is
independent of the phantom chemical potential (contrary to the
requirement of \cite{Lima:2008qx}) and the change in the sign of
$\dot{M}$ is consistent with the change in the $T$-sign
\cite{phantomtherm}. Although smooth at the  $-1$-crossing, such a
change in the variation rate of a black hole mass, and thus of the
total  mass in black holes in the universe, could leave observable
imprints in the celestial orbits.\\

\paragraph*{{\bf{Acknowledgements:}}}
E. N. Saridakis wishes to thank Institut de Physique Th\'eorique,
CEA, for the hospitality during the preparation of the present
work.


\begin{thebibliography}{0}




\bibitem{observ}
A.G. Riess {\it et al.} [Supernova Search Team Collaboration],
Astron. J. {\bf 116}, 1009 (1998);
 S.
Perlmutter {\it{et al.}} [Supernova Cosmology Project
Collaboration], Astrophys. J. {\bf 517}, 565 (1999);
 D. N. Spergel {\it{et al.}}, Astrophys.
J. Suppl. {\bf 148}, 175 (2003); S. W. Allen, {\it{et al.}}, Mon.
Not. Roy. Astron. Soc. {\bf 353}, 457 (2004).


\bibitem{ordishov}
 P. Bin\'{e}truy, C. Deffayet, D. Langlois, Nucl. Phys. B {\bf565}, 269 (2000);
G.R. Dvali, G. Gabadadze, M. Porrati, Phys. Lett. B {\bf485}, 208
(2000); S. Capozziello, Int. J. Mod. Phys. D {\bf11}, 483 (2002);
 S.Nojiri
and S.~D.~Odintsov, Phys. Rev. D {\bf{68}}, 123512 (2003);
%\cite{Apostolopoulos:2005ff}
  P.~S.~Apostolopoulos, N.~Brouzakis, E.~N.~Saridakis and N.~Tetradis,
  %``Mirage effects on the brane,''
  Phys.\ Rev.\  D {\bf 72}, 044013 (2005);
  %%CITATION = PHRVA,D72,044013;%%
S.Nojiri and S.~D.~Odintsov, Int. J. Geom. Meth. Mod. Phys.
{\bf{4}}, 115 (2007);
%\cite{Diakonos:2007au}
  F.~K.~Diakonos and E.~N.~Saridakis,
  %%CITATION = ARXIV:0708.3143;%%
 JCAP {\bf 0902}, 030 (2009);
%\cite{Bamba:2009ay}
  K.~Bamba and C.~Q.~Geng,
  %``Thermodynamics in $F(R)$ gravity with phantom crossing,''
  arXiv:0901.1509 [hep-th].
  %%CITATION = ARXIV:0901.1509;%%


\bibitem{quint}
B.~Ratra and P.~J.~E.~Peebles, Phys.\ Rev.\ D {\bf 37}, 3406
(1988); C.~Wetterich, Nucl.\ Phys.\ B {\bf 302}, 668 (1988);
A.~R.~Liddle and R.~J.~Scherrer, Phys.\ Rev.\ D {\bf 59}, 023509
(1999); I.~Zlatev, L.~M.~Wang and P.~J.~Steinhardt, Phys.\ Rev.\
Lett.\ {\bf 82}, 896 (1999);
%\cite{Podolsky:2002dp}
  D.~Podolsky,
  Astron.\ Lett.\  {\bf 28}, 434 (2002);
  %%CITATION = ALETE,28,434;%%
 Z.~K.~Guo, N.~Ohta and Y.~Z.~Zhang,
Mod.\ Phys.\ Lett.\  A {\bf 22}, 883 (2007).

\bibitem{phant} R. R. Caldwell, Phys.
Lett. B {\bf{545}}, 23 (2002); R.~R.~Caldwell, M.~Kamionkowski and
N.~N.~Weinberg, Phys. Rev. Lett. {\bf 91}, 071301 (2003);  V. K.
Onemli and R. P. Woodard, Phys.\ Rev.\ D {\bf 70}, 107301 (2004);
%\cite{Setare:2008mb}
  M.~R.~Setare and E.~N.~Saridakis,
 JCAP {\bf 0903}, 002 (2009).
  %%CITATION = ARXIV:0811.4253;%%


\bibitem{quintom}
B.~Feng, X.~L.~Wang and X.~M.~Zhang, Phys.\ Lett.\  B {\bf 607},
35 (2005);
  %%CITATION = PHLTA,B607,35;%%
Z. K. Guo, {\it{et al.}}, Phys. Lett. B {\bf 608}, 177 (2005);
M.-Z Li, B. Feng, X.-M Zhang, JCAP,  {\bf0512}, 002 (2005); W.
Zhao and Y. Zhang, Phys. Rev. D {\bf73}, 123509 (2006);
  %%CITATION = JCAPA,0809,026;%%
  %\cite{Setare:2008pz}
  M.~R.~Setare and E.~N.~Saridakis,
  %``Coupled oscillators as models of quintom dark energy,''
  Phys.\ Lett.\  B {\bf 668}, 177 (2008);
  %%CITATION = PHLTA,B668,177;%%
%\cite{Setare:2008si}
  M.~R.~Setare and E.~N.~Saridakis,
  %``Quintom model with O($N$) symmetry,''
  JCAP {\bf 0809}, 026 (2008).
  %%CITATION = JCAPA,0809,026;%%



\bibitem{phantomtherm}
  P.~F.~Gonzalez-Diaz and C.~L.~Siguenza,
  Nucl.\ Phys.\  B {\bf 697}, 363 (2004).
  %%CITATION = NUPHA,B697,363;%%






%\cite{Myung:1900km}
\bibitem{Myung:1900km}
  Y.~S.~Myung,
  %``On phantom thermodynamics with negative temperature,''
  Phys.\ Lett.\  B {\bf 671}, 216 (2009).
  %%CITATION = PHLTA,B671,216;%%





%\cite{Lima:2004wf}
\bibitem{Lima:2004wf}
  J.~A.~S.~Lima and J.~S.~Alcaniz,
  %``Thermodynamics and spectral distribution of dark energy,''
  Phys.\ Lett.\  B {\bf 600}, 191 (2004).
  %%CITATION = PHLTA,B600,191;%%


%\cite{Lima:2008uk}
\bibitem{Lima:2008uk}
  J.~A.~S.~Lima and S.~H.~Pereira,
  %``Chemical Potential and the Nature of the Dark Energy: The case of
  %phantom,''
  Phys.\ Rev.\  D {\bf 78}, 083504 (2008).
  %%CITATION = PHRVA,D78,083504;%%

%\cite{Pereira:2008af}
\bibitem{Pereira:2008af}
  S.~H.~Pereira and J.~A.~S.~Lima,
  %``On Phantom Thermodynamics,''
  Phys.\ Lett.\  B {\bf 669}, 266 (2008).
  %%CITATION = PHLTA,B669,266;%%


\bibitem{Szydlowski:2008zzaB}
  M.~Szydlowski, O.~Hrycyna and A.~Kurek,
  %``Coupling constant constraints in a nonminimally coupled phantom
  %cosmology,''
  Phys.\ Rev.\  D {\bf 77}, 027302 (2008).
  %%CITATION = PHRVA,D77,027302;%%


\bibitem{Hilbert} R. Courant and D. Hilbert, {\it{Methods of Mathematical
Physics}}, Vol. 2, John Wiley $\&$ Sons, New York (1962).


  \bibitem{Cline:2003gs}
    J.~M.~Cline, S.~Jeon and G.~D.~Moore,
    %``The phantom menaced: Constraints on low-energy effective ghosts,''
    Phys.\ Rev.\  D {\bf 70}, 043543 (2004).
    %%CITATION = PHRVA,D70,043543;%%



\bibitem{quantumphantom}
  S.~Nojiri and S.~D.~Odintsov,
  Phys.\ Lett.\  B {\bf 562}, 147 (2003);
  S.~Nojiri and S.~D.~Odintsov,
  Phys.\ Lett.\  B {\bf 571}, 1 (2003);
  %%CITATION = PHLTA,B571,1;%%
  D.~Samart and B.~Gumjudpai,
  Phys.\ Rev.\  D {\bf 76}, 043514 (2007).
  %%CITATION = PHRVA,D76,043514;%%


\bibitem{Babichev:2004yx}
  E.~Babichev, V.~Dokuchaev and Yu.~Eroshenko,
  Phys.\ Rev.\ Lett.\  {\bf 93}, 021102 (2004).
  %%CITATION = PRLTA,93,021102;%%



\bibitem{Lima:2008qx}
  J.~A.~S.~Lima, S.~H.~Pereira, J.~E.~Horvath and D.~C.~Guariento,
  [arXiv:0808.0860 [gr-qc]].
  %%CITATION = ARXIV:0808.0860;%%



\end{thebibliography}
\end{document}